\documentstyle[twocolumn,aps,epsfig]{revtex}

\newcommand{\Pbar}{\overline{P}}
\newcommand{\Qbar}{\overline{Q}}
\newcommand{\sPQ}{\sigma^2_{\scriptscriptstyle\overline{P/Q}}}

\begin{document}

\tolerance = 10000

\twocolumn[\hsize\textwidth\columnwidth\hsize\csname @twocolumnfalse\endcsname

\title{Error estimation in the histogram Monte Carlo method}
\author{M. E. J. Newman}
\address{Santa Fe Institute, 1399 Hyde Park Road, Santa Fe, New Mexico 87501}
\author{R. G. Palmer}
\address{Department of Physics, Box 90305, Duke University, Durham,
  North Carolina 27708}
\date{\today}
\maketitle

\begin{abstract}%
  We examine the sources of error in the histogram reweighting method for
  Monte Carlo data analysis.  We demonstrate that, in addition to the
  standard statistical error which has been studied elsewhere, there are
  two other sources of error, one arising through correlations in the
  reweighted samples, and one arising from the finite range of energies
  sampled by a simulation of finite length.  We demonstrate that while the
  former correction is usually negligible by comparison with statistical
  fluctuations, the latter may not be, and give criteria for judging the
  range of validity of histogram extrapolations based on the size of this
  latter correction.
\end{abstract}


]

\section{Introduction}
Monte Carlo simulations have a long and interesting history.  As a tool for
studying physical systems (rather than for performing integrals), they date
back at least as far as the pioneering work on neutron diffusion by Enrico
Fermi in the 1930s\cite{Segre80}, but Monte Carlo methods really came to
prominence in the fifties following the calculations on hard-sphere gases
and other simple systems performed by Ulam, Metropolis, von Neumann and
others using the early digital computers at Aberdeen and Los
Alamos\cite{MHR80}.  In the last three decades, with the availability of
ever-increasing amounts of computer power, the Monte Carlo method has
become one of the most important tools in the statistical physicist's
tool-box.

Although the name Monte Carlo covers a multitude of different ideas and
techniques, we concentrate in this paper on the simulation of classical
models in thermal equilibrium.  All equilibrium Monte Carlo calculations
revolve around the same fundamental idea.  One generates a number of states
$i=1\ldots n$ of the system of interest and measures for each one the total
energy $E_i$ and any other quantities of interest $X_i$, $Y_i$, etc.
Normally all states $i$ are not generated equally probably, but with
varying probabilities $p_i$, a technique known as importance sampling.  The
best estimate of the thermal average $\langle X \rangle$ of a quantity $X$
is then given by
\begin{equation}
\langle X \rangle = {\sum_i X_i p_i^{-1} {\rm e}^{-\beta E_i}\over
                     \sum_i p_i^{-1} {\rm e}^{-\beta E_i}},
\label{expectation}
\end{equation}
where $\beta=(kT)^{-1}$ is the inverse temperature and $k$ is the Boltzmann
constant.  In some cases, particularly in systems which display symmetry-
or ergodicity-breaking, we may not in fact wish to calculate an average
over all states in this way\cite{Palmer82}.  For the purposes of this
paper however, we assume that we are working with ergodic systems for which
expectations of the form~(\ref{expectation}) are physically meaningful.

The most common choice by far for the probabilities $p_i$ is to make them
proportional to the Boltzmann weight of the corresponding state at the
temperature of interest
\begin{equation}
p_i \propto {\rm e}^{-\beta E_i},
\label{boltzmann}
\end{equation}
in which case Equation~(\ref{expectation}) reduces to a simple average over
the measurements $X_i$.  Many other choices have been investigated however,
including simple or uniform sampling\cite{BH92} in which $p_i$ is a
constant independent of $i$, entropic sampling\cite{Lee93} in which $p_i$
is proportional to the reciprocal of the density of states at energy $E_i$,
and $1/k$ sampling\cite{HS95} in which $p_i$ is proportional to the
reciprocal of the integrated density of states.  In the present paper we
investigate the case in which the states are sampled with probabilities
proportion to their Boltzmann weights, but at a temperature $T_0$ different
from the temperature at which we wish to calculate $\langle X \rangle$.  In
other words, we imagine performing a normal thermal Monte Carlo simulation
at a temperature $T_0$, and then ask for the best estimate of the
expectation of $X$ at a different temperature $T$.  Making the replacement
$\beta\to\beta_0$ in Equation~(\ref{boltzmann}) and substituting
into~(\ref{expectation}), we obtain
\begin{equation}
\langle X \rangle = {\sum_i X_i {\rm e}^{-(\beta-\beta_0)E_i}\over
                     \sum_i {\rm e}^{-(\beta-\beta_0)E_i}}.
\label{single}
\end{equation}
This is not a new result.  Already in 1972, Valleau and Card\cite{VC72}
pointed out that it is possible in theory to extract a value for $\langle X
\rangle$ at any temperature from the results of a single thermal Monte
Carlo simulation using an equation of this type.  Their results were
rediscovered and extended in 1988 by Ferrenberg and Swendsen\cite{FS88},
who dubbed this technique the ``single histogram method''.  The name is
something of a misnomer, since the method's application does not
necessarily involve the construction of any histograms.  Ferrenberg and
Swendsen's formulation however was in terms of histograms and, as we will
see, it is often convenient to represent the method in this way.

Defining the double
histogram $H(E,X)$ to be the number of states $i$ sampled for which $E_i=E$
and $X_i=X$, we can rewrite Equation~(\ref{single}) in the form
\begin{equation}
\langle X \rangle = {\sum_{E,X} X H(E,X)\> {\rm e}^{-(\beta-\beta_0)E}\over
                     \sum_{E,X} H(E,X)\> {\rm e}^{-(\beta-\beta_0)E}}.
\label{sh}
\end{equation}
If we define a set of weights
\begin{equation}
W(E,X) = H(E,X)\> {\rm e}^{-(\beta-\beta_0)E}
\label{eweights}
\end{equation}
then Equation~(\ref{sh}) can be rewritten as a weighted average over $X$:
\begin{equation}
\langle X \rangle = {\sum_{E,X} X W(E,X)\over\sum_{E,X} W(E,X)}.
\end{equation}
Note that $W(E,X)$ and $H(E,X)$ become equal when $\beta=\beta_0$.  In
effect, $W(E,X)$ is an estimate of the value of the histogram $H(E,X)$ at
the temperature of interest.

It is possible to write an equation similar to~(\ref{single}) for
parameters other than the temperature, allowing us to extrapolate the
results of a single simulation to other values of any external field
appearing in the Hamiltonian.  It is also straightforward to generalize the
histogram method to non-Boltzmann sampling schemes.
Here however we concentrate on the simple case described
above.

In this paper we explore the sources of error in histogram extrapolations.
The statistical errors inherent in the method have been discussed at some
length elsewhere\cite{FLS95}, and it is not our intention to reproduce
previous results here.  We focus instead on two important sources of error
which have been neglected in previous studies.  In Section~\ref{sample} we
discuss errors introduced as a result of the finite range of energies
sampled in a simulation of finite length, and show that in certain
temperature regimes this, and not statistical fluctuation, is the dominant
source of error.  In Section~\ref{distribution} we discuss errors
introduced by the correlation between fluctuations in the numerator and
denominator of Equation~(\ref{single}).  In Section~\ref{statistical} we
discuss corrections to the normal expression for the statistical errors
arising from the previous analysis and show that to leading order these
corrections are negligible.  In Section~\ref{concs} we give our
conclusions.

\section{Finite sample size errors}
\label{sample}
Suppose that we perform a single Monte Carlo simulation at temperature
$T_0$ on some system of interest, and that this simulation samples $n$
states of the system at intervals of $\tau_s$ Monte Carlo steps.  We assume
in this paper that $\tau_s$ is much greater than the correlation time
$\tau$ of the simulation algorithm used (also measured in Monte Carlo
steps) so that the states may be considered to be statistically
independent.  More generally, if $\tau_s$ and $\tau$ are comparable, then
the variance in a measured quantity is increased by a factor of
$1+2\tau/\tau_s$ over its value for uncorrelated samples\cite{MKB73}.  All
the results given in this paper can be generalized to this case in a
straightforward manner; see Ref.~\onlinecite{FLS95} for a thorough
exploration of this issue.

\begin{figure}[t]
\begin{center}
\psfig{figure=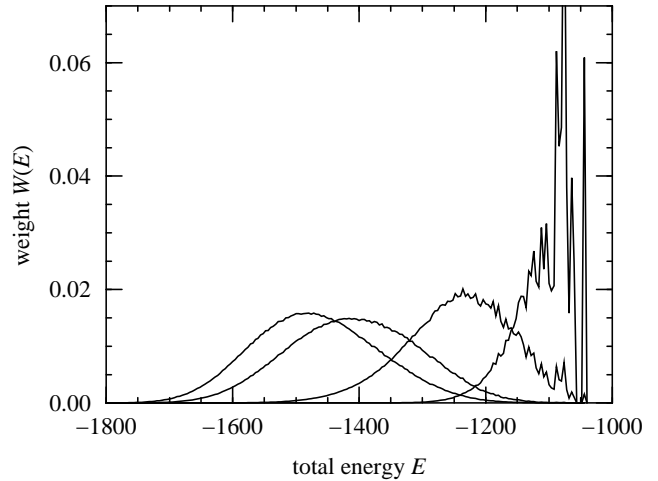,width=8cm}
\end{center}
\caption{The weight function $W(E)$ for a $32\times32$ Ising ferromagnet on
  a square lattice in two dimensions, calculated at four different
  temperatures from a single simulation at the critical temperature
  $T_c=2.269$ of the infinite system.  The curves shown are (left to right)
  for $T=T_c$, $2.3$, $2.4$, and $2.6$.
\label{weights}}
\end{figure}

In the limit of an infinite number of independent samples, $n\to\infty$,
Equation~(\ref{single}) is exact and correctly gives the value of $\langle
X \rangle$ at all temperatures.  In practice, however, $n$ is always
finite, and this limits the range over which the extrapolation is valid.
In Figure~\ref{weights} we show an example of the use of the single
histogram method to calculate the internal energy of a two-dimensional
Ising model in zero field.  The case of the internal energy is particularly
simple, since the weight function $W(E,X)$ reduces in this case to a
function $W(E)$ of a single variable $E$, the energy of the states sampled
in the simulation.  The figure shows the calculated value of this function
for a variety of different temperatures at distances increasingly far from
the temperature $T_0$ of the original simulation.  For small deviations
from $T_0$ the calculated value of $W(E)$ is a good approximation to the
histogram $H(E)$ which would be generated by a simulation performed at
temperature $T$.  However as $T$ strays farther from $T_0$, the value of
$W(E)$ becomes an increasingly poor representation of the correct
histogram, as can be seen in the figure.  The source of this problem is
clear: a finite-$n$ Monte Carlo simulation samples energies in only a
rather narrow range around the value $U(T_0)$ of the equilibrium internal
energy of the system at $T_0$.  Extrapolation of the results to
temperatures $T$ for which the true histogram $H(E)$ would possess
significant contributions at energies outside this range is therefore
guaranteed to give poor results.  In the particular case of the internal
energy, it is clear that if the highest energy sampled by our simulation is
$E_+$, then no reweighting of our histogram can ever produce an estimate of
$U(T) \equiv \langle E \rangle$ greater than $E_+$, regardless of the true
value.

The usual rule of thumb for estimating the range of validity of the
extrapolation is to require that the mean of the reweighted distribution
$W(E)$, which is just the internal energy $U(T)$, should be less than
$\sigma_E$ away from the mean $U(T_0)$ of the histogram $H(E)$, where
$\sigma_E$ is the standard deviation of $H(E)$.  Since $\sigma_E$ is
related to the specific heat $C$ at $T_0$ according to $C(T_0) =
k\beta_0^2\sigma_E^2$, we can also express this condition in terms of
$C(T_0)$ as
\begin{equation}
[U(T)-U(T_0)]^2 < kT_0^2 C(T_0).
\label{condition1}
\end{equation}

Equation~(\ref{condition1}) can be simplified further if we make the
derivative approximation
\begin{equation}
U(T) - U(T_0) \simeq (T - T_0) {{\rm d}U\over{\rm d}T}\bigg|_{T_0}
                    = \Delta T\>C(T_0),
\label{derivapp}
\end{equation}
where $\Delta T \equiv T - T_0$ is the temperature range over which we are
extrapolating.  Employing this approximation, our condition becomes
\begin{equation}
\biggl[{\Delta T\over T_0}\biggr]^2 < {k\over C(T_0)}.
\label{condition2}
\end{equation}
This condition is intuitively easy to understand and in most cases is a
reasonable guide for applying the histogram method.  However, as we will
demonstrate, the actual range of validity of the method can deviate
arbitrarily far from the value of $\Delta T$ given by
Equation~(\ref{condition2}), depending on the number $n$ of samples
generated by the Monte Carlo simulation.

We now construct a more accurate criterion for the extrapolation range.
The basic idea is to make an estimate of the energy $E_+$ above which there
are no samples, and then to approximate the error introduced into our
extrapolation by assuming that the histogram is accurate up to $E_+$, and
contains no samples thereafter.  We do the same for the lower limit $E_-$
of the histogram.  A variation on this idea would be to restrict the
extrapolation to a range of energies such that some prescribed fraction of
the samples in the histogram fall within that range.  However, since the
tails of the histogram typically decay exponentially or faster, these two
approaches give approximately the same results.

Consider the ideal histogram $\overline{H(E)}$, which we define to be the
value of the histogram $H(E)$ averaged, bin by bin, over an infinite number
of simulations which generate $n$ samples each.  We then approximate the
histogram resulting from a single simulation by
\begin{equation}
H(E) = \biggl\lbrace\begin{array}{ll}
       (n/n')\overline{H(E)}\quad & \mbox{if $E_-<E<E_+$}\\
       0                          & \mbox{otherwise.}
       \end{array}
\label{HE}
\end{equation}
The factor $n/n'$, where $n' = \int_{E_-}^{E_+}\overline{H(E)}\,{\rm d}E$,
is a normalizing factor which ensures that the integral of $H(E)$ over $E$
is correctly equal to $n$.  The values of $E_+$ and $E_-$ are defined
naturally by
\begin{equation}
\overline{H(E_\pm)} = a,
\label{defsa}
\end{equation}
where $a$ is a constant of order unity.

Making this approximation, the extrapolated internal energy $U(T)$ can be
written as
\begin{eqnarray}
\Delta U &=& \overline{U(T)} - U(T)\nonumber\\
     &=& {\int E\,{\rm e}^{(\beta-\beta_0)E}
          \,\overline{H(E)}\> {\rm d} E
         \over
          \int {\rm e}^{(\beta-\beta_0)E}
          \,\overline{H(E)}\> {\rm d} E
         }
         -
         {\int E\,{\rm e}^{(\beta-\beta_0)E}
          H(E)\> {\rm d} E
         \over
          \int {\rm e}^{(\beta-\beta_0)E}
          H(E)\> {\rm d} E
         }\nonumber\\
     &=& {\partial\over\partial\beta} \log
         {
          \int_{E_-}^{E_+}
          {\rm e}^{-(\beta-\beta_0)E} \,\overline{H(E)}\> {\rm d} E
         \over
          \int_{-\infty}^{\infty}
          {\rm e}^{-(\beta-\beta_0)E} \,\overline{H(E)}\> {\rm d} E
         }.
\label{deltau1}
\end{eqnarray}

In order to proceed we make a Gaussian approximation for $\overline{H(E)}$:
\begin{equation}
\overline{H(E)} = {n\over\sqrt{2\pi\sigma_E^2}}\,\exp \biggl(-{[E
  -U(T_0)]^2\over2\sigma_E^2}\biggr).
\label{gaussian}
\end{equation}
This assumption is an excellent guide for the behavior of most systems at
temperatures well above $T=0$.  For instance, in the Ising system of
Figure~\ref{weights} it gives $\log\overline{H(E)}$ within a few percent
over more than a hundred orders of magnitude of $\overline{H(E)}$.

Using Equation~(\ref{gaussian}) and another derivative approximation:
\begin{equation}
(\beta-\beta_0)\,\sigma_E^2 = -(\beta-\beta_0) {{\rm d}U\over{\rm
    d}\beta}\Big|_{\beta_0} \simeq U(T_0) - U(T),
\end{equation}
we complete the square to obtain
\begin{eqnarray}
&\overline{H(E)}& {\rm e}^{-(\beta-\beta_0)E}\simeq\nonumber\\
   & & {n\over\sqrt{2\pi\sigma_E^2}}\, f(\beta)\,
       \exp \biggl(-{[E-U(T)]^2\over2\sigma_E^2}\biggr),
\label{quadapp}
\end{eqnarray}
where
\begin{equation}
f(\beta) = \exp\biggl({U^2(T)-U^2(T_0)\over2\sigma_E^2}\biggr)
\label{fofbeta}
\end{equation}
is a shorthand for all the terms in the exponential which depend on $\beta$
but not on $E$.  Substituting Equation~(\ref{quadapp}) into~(\ref{deltau1})
and performing the integral leads to
\begin{eqnarray}
\Delta U &=& {\partial\over\partial\beta}
             \log \biggl[ \mbox{$\frac12$} \mathop{\rm erf}
             \biggl({E-U(T)\over\sqrt{2}\sigma_E}
             \biggr)\biggr]_{E_-}^{E_+}\nonumber\\
         &=& {\sqrt{2\sigma_E^2\over\pi}}\,
            {\exp(-x^2_+) - \exp(-x^2_-)\over
             \mathop{\rm erf}(x_+) - \mathop{\rm erf}(x_-)},
\label{deltau2}
\end{eqnarray}
where $\mathop{\rm erf}(x)={2\over\sqrt{\pi}}\int_0^x{\rm e}^{-t^2}\,{\rm d}t$
is the Gaussian error function, and
\begin{equation}
x_{\pm} \equiv {E_{\pm} - U(T)\over\sqrt{2}\sigma_E} =
  \pm\sqrt{\log{n\over\sqrt{2\pi}a\sigma_E}} -
  {\sigma_E\Delta T\over\sqrt{2}kT_0^2},
\label{epm}
\end{equation}
using Equations~(\ref{derivapp}), (\ref{defsa}) and ~(\ref{gaussian}).

Between them, Equations~(\ref{deltau2}) and~(\ref{epm}) give us an
estimate of the deviation of the extrapolation of $U$ from its true value
as a function of the number of samples $n$ and the temperature range
$\Delta T$ over which we extrapolate.

\begin{figure}[t]
\begin{center}
\psfig{figure=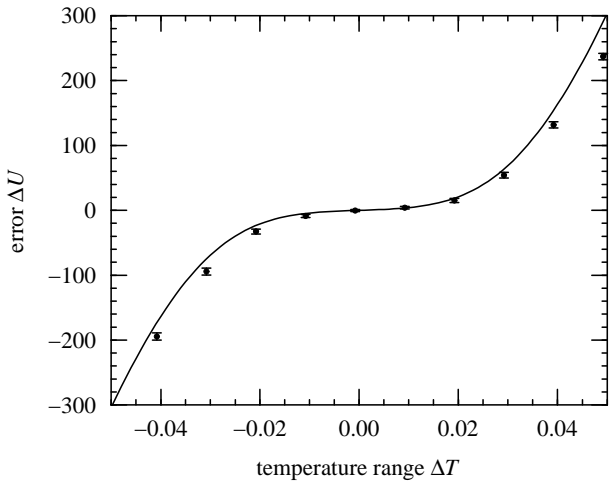,width=8cm}
\end{center}
\caption{The difference $\Delta U$ between the true internal energy of a
  $100\times100$ Ising ferromagnet and an extrapolation using
  Equation~(\ref{single}) of the same quantity from simulations with
  $n=100$ samples performed at a single temperature $T_0 = 2.269$.  The
  line is a fit using Equations~(\ref{deltau2}) and~(\ref{epm}).  Energies
  are in units of the coupling constant $J$, and may be compared to $U(T_0)
  = -1.4\times 10^4$.
\label{ducomp}}
\end{figure}

As a test of this calculation we have plotted in Figure~\ref{ducomp} the
value of $\Delta U$ measured in simulations of a $100\times100$ Ising model
on a square lattice in two dimensions.  The data points with error bars
show the difference between the true internal energy (obtained from further
independent simulations) and those calculated via Equation~(\ref{single})
from simulations with $n=100$ samples at temperature $T_0 = 2.269$ (the
critical temperature of the infinite system).  These points are averaged
over 1000 repetitions of the simulation at $T_c$.  The solid line is from
Equations~(\ref{deltau2}) and~(\ref{epm}) with the constant $a$ chosen so
as to best fit the data.  As the figure shows, the agreement between the
two is good.

In a typical Monte Carlo calculation we want to know the range of
temperature $\Delta T$ over which we can extrapolate from a single
histogram to a given degree of accuracy $\Delta U$ as a function of the
sample size $n$.  In the regime where $U(T)$ approaches either of the
limits $E_-$ or $E_+$, one or other of the terms on the top and bottom of
Equation~(\ref{deltau2}) becomes a constant (either zero or one) and the
variation in $\Delta U$ resides entirely in the remaining terms.  In this
case a line of constant $\Delta U$ is also a line of constant $x_+$ or
$x_-$ (for $\Delta T$ positive or negative respectively) which means that
\begin{equation}
\pm\sqrt{\log{n\over\sqrt{2\pi}\sigma_Ea}} -
  {\sigma_E\Delta T\over\sqrt{2}kT_0^2} = b,
\label{criterion}
\end{equation}
with the value of the constant $b$ depending on the size of error $\Delta U$ we
are willing to live with.  Thus, for given $\Delta U$, the temperature
range $\Delta T$ over which the extrapolation is valid increases at most
logarithmically with increasing sample size $n$.

\begin{figure}[t]
\begin{center}
\psfig{figure=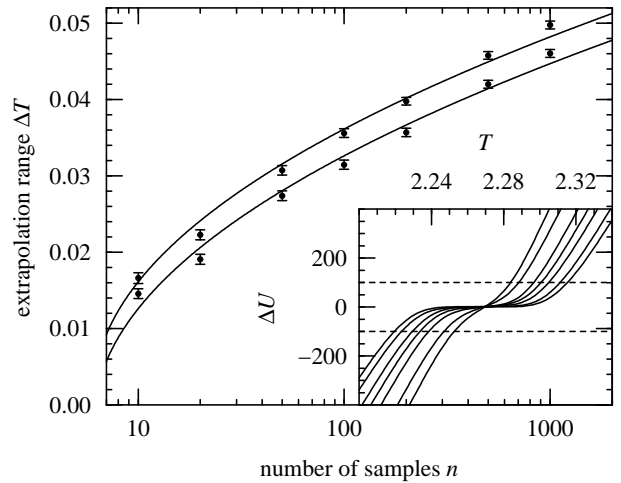,width=8cm}
\end{center}
\caption{Inset: the difference between the true and extrapolated internal
  energies of a $100\times100$ Ising ferromagnet for a variety of different
  sample sizes $n$.  Main figure: the range $\Delta T$ over which the
  extrapolation is accurate to $\pm100$, as a function of $n$.  The points
  are the values from the simulations shown in the inset and the two solid
  lines are Equation~(\ref{criterion}), taking the $+$ and $-$ signs
  separately.  The upper curve and points are for positive $\Delta T$,
  the lower ones for negative $\Delta T$.
\label{crit}}
\end{figure}

In Figure~\ref{crit} we demonstrate this formula for the $100\times100$
two-dimensional Ising model.  The inset shows extrapolations from the
critical temperature of the infinite system for sample sizes $n=10$, $20$,
$50$, $100$, $200$, $500$, and $1000$, using Equation~(\ref{single}).  The
errors in these results are comparable to the widths of the lines.  The
dashed lines show an arbitrarily-chosen deviation of $\Delta U=\pm 100$
from the true value as our limit of acceptable accuracy---a relative error
of about 0.7\%.  The intersections of the solid curves and dashed lines
give the ranges $\Delta T$ over which simulations with different $n$ give
acceptable results.  The main figure shows these ranges as points with
error bars, the upper points corresponding to values $\Delta T>0$
(i.e.,~extrapolation above $T_0$), the lower ones to $\Delta T<0$.  The
solid lines are Equation~(\ref{criterion}) with the constants $a$ and $b$
chosen by a least squares fit to the data.  As the figure shows, simulation
and theory are in good agreement.

As an example of the use of Equation~(\ref{criterion}), consider the
results of M\"unger and Novotny\cite{MN91} who performed an extensive
numerical study of the accuracy of the single histogram extrapolation
method for the case of the $q=3$ Potts ferromagnet in two dimensions.  They
concluded that the values of the specific heat predicted by the method show
systematic deviations from the true values, and presented evidence
indicating that the size of these deviations decrease with increasing $n$.
In fact, a simple application of Equations~(\ref{condition2})
and~(\ref{criterion}) reveals immediately what the problem is.  For the
parameter values and sample sizes used in their calculations, the range
over which they attempt to extrapolate satisfies the simple
criterion~(\ref{condition2}), but falls outside the bounds of accuracy set
by~(\ref{criterion}).

M\"unger and Novotny deliberately performed simulations with small values
of $n$ in order to investigate the inaccuracies of the histogram method.
However, in normal use, the method is applied to simulations with large
$n$, and in the region close to $T_0$ where the deviation $\Delta U$ is
small.  We can characterize this regime as one in which $|x_\pm|\gg 1$, in
which case the value of the denominator in Equation~(\ref{deltau2}) is
close to~2 and the primary variation in $\Delta U$ comes from the Gaussians
in the numerator:
\begin{equation}
\Delta U \simeq \sqrt{\sigma_E^2\over2\pi}\,
                \bigl[\exp\bigl(-x_+^2\bigr) - \exp\bigl(-x_-^2\bigr)\bigr].
\label{deltau3}
\end{equation}
Since $E_+$ and $E_-$ are symmetrically distributed about $U(T_0)$, we have
$x_+(T_0) = -x_-(T_0)$, and the two terms cancel to give $\Delta U=0$ at
$T=T_0$, as expected.  The leading term in the expansion of $\Delta U$
about this point is linear in $\Delta T$ with coefficient
\begin{equation}
{\partial\Delta U\over\partial T}\bigg|_{T_0} =
  {2a\beta_0^2\sigma_E^3\over n} \sqrt{\log {n^2\over2\pi\sigma_E^2 a^2}}.
\label{scaling}
\end{equation}
Thus $\Delta U$ tends to zero roughly as $1/n$ to leading order, and the
higher order terms vanish faster than this.  As we will see in
Section~\ref{statistical}, the statistical errors in extrapolated
quantities fall off in the normal $1/\sqrt{n}$ fashion, so that in the
region close to $T_0$, finite sample size errors always become negligible
for sufficiently large $n$.  

On the other hand, when we get far away from $T_0$, the extrapolated value
of $U$ becomes roughly equal to $E_+$ or $E_-$ (depending on the direction
in which we extrapolate) and hence approximately independent of $n$, since
$E_\pm$ only varies slowly with $n$.  Thus the error $\Delta U$ is
approximately $n$-independent in this regime and dominates over statistical
errors for sufficiently large $n$.  The point of crossover between the two
regimes is given by Equation~(\ref{criterion}).

A similar argument can be made for the extrapolation of quantities other
than the energy.  The limiting extrapolated values of any quantity $Y$ are
set by the values $Y_\pm$ corresponding to the highest and lowest energies
sampled in the simulation, and since these energies are approximately
$n$-independent, so normally will $Y_\pm$ be.  Thus
Equation~(\ref{criterion}) tells us for any quantity $Y$ the point of
crossover at which errors due to the finite number of samples in the
histogram become the dominant source of inaccuracy in the histogram method.

\section{Distribution errors}
\label{distribution}
There is another source of systematic error in the estimates given by the
single histogram method which has not, to our knowledge, been remarked upon
before.  Even ignoring the corrections discussed in the last section, which
were due to the imperfect sampling of the histogram $H(E)$,
Equation~(\ref{single}) is not in fact a correct expression for the best
estimate of $\langle X \rangle$ for any finite $n$.  To understand this,
consider again the hypothetical situation in which we perform a large
number $N$ of simulations of the system of interest, each one generating
$n$ statistically independent samples drawn from the Boltzmann distribution
at $T_0$.  For each one we calculate an estimate
\begin{equation}
\langle X \rangle_i = {\sum_j X_{ij} {\rm e}^{-(\beta-\beta_0)E_{ij}}\over
                      \sum_j {\rm e}^{-(\beta-\beta_0)E_{ij}}}
                    = {P_i\over Q_i},
\label{manysingle}
\end{equation}
where $i=1\ldots N$ labels the different simulations and $X_{ij}$ is the
value of $X$ in the $j$th state sampled by the $i$th simulation.  The new
quantities $P$ and $Q$ will provide a convenient shorthand for the
numerator and denominator of this equation.

Now we want to compute the best estimate of $\langle X \rangle$ over
all $N$ simulations.  Since the samples in each simulation were drawn from
the same distribution, we can just as well regard them all as being one
large set of samples of size $nN$ drawn from a single simulation, in which
case it is clear that in the limit of large $N$ the correct answer for
$\langle X \rangle$ is
\begin{equation}
\langle X \rangle = {\sum_{ij} X_{ij} {\rm e}^{-(\beta-\beta_0)E_{ij}}\over
                     \sum_{ij} {\rm e}^{-(\beta-\beta_0)E_{ij}}}
                  = {\,\Pbar\,\over\Qbar},
\label{avsingle}
\end{equation}
where $\Pbar$ and $\Qbar$ indicate the averages of $P_i$ and
$Q_i$ over all $N$ simulations.  (We use the barred notation to avoid
confusion with the notation $\langle X \rangle$ for thermal expectation
values.)  This equation indicates that the best estimate of $\langle X
\rangle$ is calculated by separately averaging the numerator and
denominator of Equation~(\ref{manysingle}) over our many simulations.  In
practice, one does not perform many simulations; one performs only one
simulation with finite $n$ and then calculates the ratio $P/Q$ for that one
simulation.  The mean value of this ratio however is not the same as the
ratio of the means, Equation~(\ref{avsingle}), which gives the correct
answer.  This difference leads to a systematic error in the predictions of
the single histogram method for finite sample sizes.  In this section we
calculate the size of this error.

Consider the double Taylor expansion of the quantity $P/Q$ around
$\Pbar/\Qbar$:
\begin{eqnarray}
{P\over Q} &=& {\Pbar\over\,\Qbar\,}
               + (P-\Pbar) {1\over\,\Qbar\,}
               - (Q-\Qbar) {\Pbar\over\Qbar^2}\nonumber\\
           & & + (Q-\Qbar)^2 {\Pbar\over\Qbar^3}
               - (P-\Pbar) (Q-\Qbar) {1\over\Qbar^2}
               + \ldots
\end{eqnarray}
Taking the average of both sides over many repetitions of the simulation,
the linear terms vanish and to leading order we are left with
\begin{equation}
\overline{P/Q} =
                {\Pbar\over\,\Qbar\,} \biggl[ 1 +
                   {\sigma_Q^2\over\Qbar^2}
                   - {\mathop{\rm cov}(P,Q)\over\Pbar\,\Qbar} \biggr],
\label{correction}
\end{equation}
where $\sigma_Q^2$ is the variance of $Q$ over simulations $i$ and
$\mathop{\rm cov}(P,Q)$ is the covariance of $P$ and $Q$.  Thus the mean
value of the quantity $P/Q$, which is the quantity measured in our Monte
Carlo calculations, differs from the true value of $\langle X \rangle =
\Pbar/\Qbar$ by the factor enclosed in the square brackets $[\ldots]$.  One
should take this factor into account in order to correctly calculate the
extrapolation of a quantity.

Given that in a typical situation we only perform one simulation of our
system, what is the best estimate we can make of this factor from our Monte
Carlo results?  Clearly the best estimates of $\Pbar$ and $\Qbar$ are
simply the values of $P$ and $Q$ measured in the simulation: $\Pbar = P$,
$\Qbar = Q$.  The best estimates of the variance and covariance terms are
\begin{equation}
\sigma_Q^2 = {1\over n-1} \biggl\lbrace
               \sum_j {\rm e}^{-2(\beta-\beta_0)E_j}
               - \Bigl[ \sum_j {\rm e}^{-(\beta-\beta_0)E_j}\Bigr]^2
               \biggr\rbrace,
\label{sigmaq}
\end{equation}
and
\begin{eqnarray}
\mathop{\rm cov}(P,Q) &=& {1\over n-1} \Bigl\lbrace
               \sum_j X_j {\rm e}^{-2(\beta-\beta_0)E_j}\nonumber\\
           & & - \sum_j X_j {\rm e}^{-(\beta-\beta_0)E_j}
               \sum_j {\rm e}^{-(\beta-\beta_0)E_j} \Bigr\rbrace.
\label{covariance}
\end{eqnarray}
Substituting these into Equation~(\ref{correction}) we see that the
correction term scales as $1/n$ with sample size.   But, as shown below,
statistical errors scale as $1/\sqrt{n}$ and therefore
dominate for large $n$.  Thus it should be safe to ignore errors of the
type described by Equation~(\ref{correction}) for simulations of sufficient
length.

\section{Statistical errors}
\label{statistical}
The third and final source of error which we consider is statistical
fluctuation in the extrapolation due to the essential random nature of a
Monte Carlo simulation.  We can calculate the variance
$\sPQ$ of the quantity
$\overline{P/Q}$ by a technique similar to that used to derive
Equation~(\ref{correction}); we perform a Taylor expansion of
$\overline{P^2/Q^2}$ about $\Pbar/\Qbar$ and take the average over many
simulations.  Then we calculate the variance as
$\sPQ = \overline{P^2/Q^2}-\overline{P/Q}^2$.
The variance $\sigma^2_X$ of the best estimate
of $\langle X \rangle$ is then $\sPQ$ times the
square of the
correction factor in Equation~(\ref{correction}).  To leading order this
gives
\begin{equation}
{\sigma_X^2\over\langle X \rangle^2} = 
  {\sigma_P^2\over\Pbar^2} + {\sigma_Q^2\over\Qbar^2}
  - 2 {\mathop{\rm cov}(P,Q)\over\Pbar\,\Qbar}.
\label{statvar}
\end{equation}
This expression is identical to that given by
Ferrenberg~{\it{}et~al.}\cite{FLS95}, for the error on the uncorrected
estimate $\overline{P/Q}$.

Using Equations~(\ref{sigmaq}) and~(\ref{covariance}), along with the
obvious extension
\begin{eqnarray}
\sigma_P^2 = {1\over n-1} \biggl\lbrace
               \sum_j &X_j^2& {\rm e}^{-2(\beta-\beta_0)E_j}\nonumber\\
               &-& \Bigl[ \sum_j X_j {\rm e}^{-(\beta-\beta_0)E_j}\Bigr]^2
               \biggr\rbrace,
\label{sigmap}
\end{eqnarray}
it is clear that $\sigma_X^2$ scales as $1/n$, and hence that $\sigma_X$
scales as $1/\sqrt{n}$, as claimed earlier.  This is a slower scaling than
the $1/n$ of the previous section, but still much better than the
approximately constant value of the finite sample size error of
Section~\ref{sample} for large extrapolation range $\Delta T$.  This means
that we must use an equation such as~(\ref{criterion}) to decide which of
these two latter sources of error is the dominant one under given
circumstances.

\section{Conclusions}
\label{concs}
In this paper we have examined in detail the sources of error in the Monte
Carlo extrapolation method known as the single histogram method.  We have
discussed three sources of error: finite sample size errors, systematic
errors due to the approximations made in the calculation of the
extrapolation, and finally statistical errors.  The first two of these have
not, to our knowledge been discussed previously, and in particular we find
that the finite sample size errors are, under commonly encountered
conditions, significantly larger than either of the other sources of error.

\section*{Acknowledgements}
The authors would like to thank Gerard Barkema and Catherine Macken for
useful discussions.  This research was funded in part by the Santa Fe
Institute and DARPA under grant number ONR N00014--95--1--0975.

\end{document}